\documentclass[12pt]{article}
\makeatletter
\newcommand{\singlespacing}{\let\CS=\@currsize\renewcommand{\baselinestretch}{1}\tiny\CS}
\usepackage{epsfig}
\usepackage{graphicx}
\usepackage{verbatim}   
\usepackage{color}      
\usepackage{amsfonts}
\usepackage{epstopdf}
\usepackage{amsmath}
\newtheorem{theorem}{Theorem}
\newtheorem{lemma}{Lemma}

\newtheorem{corollary}{Corollary}
\usepackage{subcaption}
\usepackage{float}
\begin{document}
\baselineskip=24pt
\parskip = 10pt
\def \qed {\hfill \vrule height7pt width 5pt depth 0pt}
\newcommand{\ve}[1]{\mbox{\boldmath$#1$}}
\newcommand{\IR}{\mbox{$I\!\!R$}}
\newcommand{\1}{\Rightarrow}
\newcommand{\bs}{\baselineskip}
\newcommand{\esp}{\end{sloppypar}}
\newcommand{\beanno}{\begin{eqnarray*}}
\newcommand{\inp}[2]{\left( {#1} ,\,{#2} \right)}
\newcommand{\eeanno}{\end{eqnarray*}}
\newcommand{\bea}{\begin{eqnarray}}
\newcommand{\eea}{\end{eqnarray}}
\newcommand{\ba}{\begin{array}}
\newcommand{\ea}{\end{array}}
\newcommand{\nno}{\nonumber}
\newcommand{\dou}{\partial}
\newcommand{\bc}{\begin{center}}
\newcommand{\ec}{\end{center}}
\newcommand{\2}{\subseteq}
\newcommand{\cl}{\centerline}
\newcommand{\ds}{\displaystyle}
\newcommand{\mr}{\mathbb{R}}
\newcommand{\mn}{\mathbb{N}}
\def\refhg{\hangindent=20pt\hangafter=1}
\def\refmark{\par\vskip 2.50mm\noindent\refhg}

\def\a{\alpha}
\def\b{\beta}
\def\d{\delta}
\def\l{\lambda}
\def\g{\gamma}
\def\Ga{\Gamma}
\def\th{\theta}

\newpage



\title{\sc Interval Estimation of the Unknown Exponential Parameter Based on Time Truncated Data}
\author{\sc Arnab Koley$^1$  \& Debasis Kundu$^2$ }

\date{}
\maketitle

\begin{abstract}
In this paper we consider the statistical inference of the unknown parameter of an exponential distribution based on the
time truncated data.  The time truncated data occurs quite often in the reliability analysis for type-I or hybrid censoring
cases.  All the results available today are based on the conditional argument that at least one failure occurs during
the experiment.  In this paper we provide some inferential results based on the unconditional argument.  We extend the results for some
two-parameter distributions also.
\end{abstract}



\noindent $^1$ Department of Mathematics and Statistics, Indian Institute of
Technology Kanpur, Pin 208016, India. E-mail: arnab@iitk.ac.in.  

\noindent $^2$ Department of Mathematics and Statistics, Indian Institute of
Technology Kanpur, Pin 208016, India. Corresponding author, E-mail: kundu@iitk.ac.in, Phone no. 91-512-2597141, Fax no. 91-512-2597500.




\section{\sc Introduction}

Let $X_1, \ldots, X_n$ be a random sample from an exponential distribution with parameter $\lambda$, then it has the following 
probability density function (PDF)
\begin{equation}
f(x; \lambda) = \begin{cases}
\lambda e^{-\lambda x}, & \text{if $x \ge 0$}, \\
0, & \text{if $x<0$.}  \label{exp-l}
\end{cases}
\end{equation}
Here $\lambda \in (0, \infty)$ is the natural parameter space. Since for $\lambda$ = 0, $f(x; \lambda)$ as defined in  
(\ref{exp-l}) is not a proper PDF, we are not including the point 0 in the parameter space.  Suppose $n$ items are tested and their ordered failure times are denoted by $x_{1:n} < x_{2:n} < \ldots <
x_{n:n}$.  If the experiment is stopped at a prefixed time $T$ then it results in a simple type-I censoring case. Let there be $D$ failures in $[0,T]$. Then $x_{1:n} < x_{2:n} < \ldots < x_{D:n} \leq T$ and 
$T < x_{D+1:n} < \ldots x_{n:n}$, although $x_{D+1:n}, \ldots, x_{n:n}$ are not observed.  Here $D$ is a random variable that can take the values 
$0, 1, \ldots, n$.  The main aim of this note is to draw inference on $\lambda$, based on $D$ observations.

This is an old problem and Bartholomew (1963) was the first to consider it.  He considered the following form of the exponential PDF;
\begin{equation}
f(x; \theta) = \begin{cases}
\frac{1}{\theta} e^{-\frac{x}{\theta}}, & \text{if $x \geq 0$},\\
0, & \text{if $x<0$}. \label{exp-t}
\end{cases}
\end{equation}
He considered $[0,\infty)$ as the parameter space of $\theta$.  In this case the maximum likelihood estimator (MLE) of
$\theta$ does not exist when $D = 0$.  Hence all the inferences related to $\theta$ are based on the conditional argument 
that $D \ge 1$.  Later on a series of papers starting with Chen and Bhattacharyya (1988) and then continuing with Gupta and Kundu (1998) and
Childs et al. (2003) considered type-I hybrid censoring case for the model (\ref{exp-t}), and obtained the exact inference
on $\theta$ based on the conditional argument that $D \ge 1$.

In case of type-I censoring, $\ds P(D = 0) = exp(-n\lambda T)$ and this probability can be 
quite high for small value of $n\lambda T$.  The natural question here is whether it is possible to draw any inference on $\lambda$, when $D = 0$.  
The second aim of the study is to determine if there exists any significant difference between the conditional and unconditional inference.  For
example, in this paper it has been shown that it is possible to construct an exact 100$(1-\alpha)\%$ confidence interval of $\lambda$ even when $D$ = 0.  Following the approach of Chen and Bhattacharyya (1998), 
an exact 100$(1-\alpha)\%$ confidence interval of $\lambda$ also can be obtained based on the conditional MLE of $\lambda$, 
conditioning on the event that $D \ge 1$.  The 
question is whether the lengths of the confidence intervals based on the two different approaches are significantly different
or not. We perform some simulation experiments to compare the confidence intervals based on the two different 
methods.  We further obtain the Bayes estimate and the associated credible interval of the unknown parameter based on the 
non-informative prior.  Finally the results are extended for the 
two-parameter exponential,  Weibull and generalized exponential distributions also.

\enlargethispage{0.75 in}
Rest of the paper is organized as follows.  In Section 2, we provide two different constructions of confidence intervals and the Bayesian credible intervals.  The simulation results for confidence and credible intervals of the parameter $\lambda$ are provided in Section 3.  In Section 4,
we extend the results for the two parameter exponential, Weibull and generalized exponential 
distributions, and finally we conclude the paper in Section 5.

\section{\sc Construction of Confidence and Credible Intervals}
In this section we proceed to construct the confidence and credible intervals of the parameter of interest $\lambda$, 
based on a new estimator of $\lambda$ and the posterior distribution of $\lambda$, respectively.
 
\subsection{\sc Confidence interval (CI)}
Based on the observations $x_{1:n} < \ldots < x_{D:n}$ from the model (\ref{exp-l}), the likelihood function becomes
\begin{equation}
L(\lambda|x_{1:n}, \ldots, x_{D:n})= \begin{cases}
e^{-n\lambda T}, & \text{if $D=0$},\\
\frac{n!}{(n-D)!} \lambda^D e^{-\lambda\big[\sum_{i=1}^D x_{i:n}+(n-D)T\big]}, & \text{if $D>0$}. \label{likelihood}
\end{cases}
\end{equation}
Note that  the MLE of $\lambda$ does not exist when $D=0$. It exists only if $D>0$ and it is given by 
$$
\widehat{\lambda}_{MLE}= \frac{D}{\sum_{i=1}^D x_{i:n} + (n-D) T}.
$$
Now based on $\ds (D, \sum_{i=1}^D x_{i:n})$, the joint sufficient statistic for $\lambda$,  
we define a new estimator of $\lambda$ for all $D \ge 0$ as follows 
\begin{equation}
\widehat{\lambda} = \frac{D}{\sum_{i=1}^D x_{i:n} + (n-D) T}.   \label{mle}
\end{equation}
Note that, $\widehat{\lambda}$ is equal to $\widehat{\lambda}_{MLE}$ only when $D>0$. Now we provide the exact distribution of 
$\widehat{\lambda}$, which will be useful in constructing an exact confidence interval of $\lambda$.
\begin{theorem}
The distribution of $\widehat{\lambda}$ for $x \ge 0$, can be written as,
\bea
P(\widehat{\lambda} \le x)  & =  &\begin{cases}
e^{-n \lambda T}, & \text{if $x=0$,}\\
e^{-n \lambda T}+\sum_{d=1}^n \sum_{k=0}^d C_{k,d} \Gamma (d, A_d(x,T_{k,d})), & \text{if $x>0$,}
\end{cases}
\label{u-dist}
\eea
\end{theorem}
where
$$
C_{k,d} = (-1)^k {n\choose{d}}{d \choose{k}} e^{-\lambda T(n-d+k)}, \ \ \ T_{k,d} = (n-d+k)T/d,
$$
\begin{equation*}
A_k(x,a) = \begin{cases}
\lambda k \left (\frac{1}{x} - a \right ),  & \text{if $x < \frac{1}{a}$},\\
0, & \text{if $x \geq \frac{1}{a}$},
\end{cases} 
\end{equation*}
and $\ds \Gamma(a,z) = \frac{1}{\Gamma (a)} \int_z^{\infty} t^{a-1} e^{-t} dt$, is the incomplete gamma function.

\noindent {\sc Proof:} For $x \geq 0$, $P(\widehat{\lambda} \leq x)$ can be written as,
\bea
P(\widehat{\lambda} \leq x)&=& P(\widehat{\lambda} \leq x,D=0)+P(\widehat{\lambda} \leq x,D>0)\nonumber \\
&=& P(\widehat{\lambda} \leq x|D=0)P(D=0)+P(\widehat{\lambda} \leq x|D>0)P(D>0)\nonumber \\
&=&\begin{cases}
e^{-n \lambda T}, & \text{if $x=0$},\\
e^{-n \lambda T}+\sum_{d=1}^n \sum_{k=0}^d C_{k,d} \Gamma (d, A_d(x,T_{k,d})), & \text{if $x>0$}.
\end{cases}
\eea
Note that the expression of $P(\widehat{\lambda} \leq x|D>0)$ can be obtained by using the moment generating function approach. Refer to Corollary 2.2 of Childs et al. (2003) or equation (8) of Gupta and
Kundu (1998) for it.  It is possible to obtain (\ref{u-dist})
in terms of the chi-square integral as in Bartholomew (1963).  The distribution of $\widehat{\lambda}$ is a mixture of  discrete 
and  continuous distributions.  The following corollary comes from Theorem 1.
\begin{corollary}
$$P(\widehat{\lambda} = 0) = e^{-n \lambda T}$$
and for $x > 0$, the PDF of $\widehat{\lambda}$ is given by
\begin{equation}
f_{\widehat{\lambda}}(x)  =  \frac{1}{x^2} \sum_{d=1}^n \sum_{k=0}^d C_{k,d} \ \ g \left ( \frac{1}{x} - T_{k,d}; \lambda d, d 
\right ), \ \ \ \ x \ge \frac{1}{nT},
\end{equation}
where 
\begin{equation}
g(x; \alpha, p) = \begin{cases}
\frac{\alpha^p}{\Gamma (p)} x^{p-1} e^{-\alpha x}, & \text{if $ x>0$},\\
0, & \text{otherwise.}
\end{cases}
\end{equation}
\end{corollary}
Now we consider the construction of an exact 100$(1-\alpha)\%$ confidence interval of $\lambda$.  We need the following lemma for further development. 
\begin{lemma}
For a fixed $b \geq 0$, $P_{\lambda}(\widehat{\lambda} \leq b)$ is a monotonically decreasing function of $\lambda$.
\end{lemma}
\noindent {\sc Proof: } The proof can be obtained along the same line as the proof of the three monotonic lemmas by Balakrishnan and Iliopoulos (2009).

\begin{figure}[h]
\includegraphics[scale=0.4]{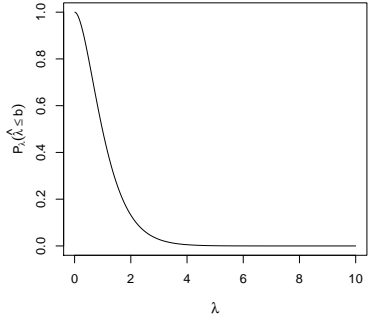}
\centering
\caption{The plot of $P_{\lambda}(\widehat{\lambda} \le b)$ as a function of $\lambda$, when $b$ is fixed.   \label{cdf-plot}}
\end{figure}

\noindent A graphical plot (Figure \ref {cdf-plot}) of $P_{\lambda}(\widehat{\lambda} \leq b)$ as a function of $\lambda$ for a fixed value of $b \geq 0$ is provided for a visual illustration.  Here we have taken $n$ = 5, $T$ = 0.5 and $b$  = 1.
We now provide an exact 100$(1-\alpha)\%$ confidence interval of $\lambda$ for different values of $D$.

\noindent \textbf{Case-I:} Construction of CI when $D=0$:\\
Since $P_{\lambda}(D = 0) = exp(-n \lambda T)$ and it is a decreasing function of $\lambda$, a one sided 100$(1-\alpha)\%$ confidence interval of $\lambda$ can be obtained as $A = \{\lambda: P_{\lambda}(D = 0) \ge 1- \alpha\}$.  
Hence, $$A = [(0, -\log(1-\alpha)/(nT)].$$\\
\noindent \textbf{Case-II:} Construction of CI when $D>0$:\\
A symmetric 
100$(1-\alpha)\%$ confidence interval, $(\lambda_L, \lambda_U)$, of $\lambda$ can be obtained by solving the following two non-linear equations
\begin{equation}
1 - \frac{\alpha}{2} = P_{\lambda_L} (\widehat{\lambda} \le \widehat{\lambda}_{obs}) \ \ \ \hbox{and} \ \ \ \
\frac{\alpha}{2} = P_{\lambda_U} (\widehat{\lambda} \le \widehat{\lambda}_{obs}).   \label{conf-1}
\end{equation}
These non-linear equations are to be solved  using standard non-linear solver \textit{viz.} Newton Rapshon, bisection method etc.
%
\subsection{\sc Credible interval (CRI)}

In this subsection we will discuss constructing a 100$(1-\alpha)\%$ credible interval of $\lambda$ based on the conjugate prior 
on $\lambda$.  It is assumed that $\lambda$ has a natural gamma prior with the shape and scale parameters 
as $a > 0$ and $b > 0$, respectively with the following PDF
\begin{equation}
\pi(\lambda)=\begin{cases}
\frac{b^a}{\Gamma a} \lambda^{a-1} e^{-\lambda b}, & \text{if $\lambda >0$},\\
0, & \text{otherwise}.
\end{cases}
\end{equation}
  Therefore the posterior density function of $\lambda$ becomes
\begin{equation}
\pi(\lambda|data) \propto \lambda^{a+D-1} e^{-\lambda\big[b + \sum_{i=1}^D x_{i:n} + (n-D)T\big]}, \ \ \  \lambda > 0.
\end{equation}
Hence the Bayes estimate of $\lambda$ under the squared error loss function becomes
\begin{equation}
\widehat{\lambda}_{Bayes} = \frac{a+D}{b + \sum_{i=1}^D x_{i:n} + (n-D)T)}.
\end{equation}
The associated 100$(1-\alpha)\%$ symmetric credible interval of $\lambda$ can be obtained as $(\lambda_{LB}, \lambda_{UB})$, 
where $\lambda_{LB}$ and $\lambda_{UB}$ can be obtained as the solutions of
\begin{equation}
\Gamma(a+D,\lambda_{LB}(b+S)) = \left (1 - \frac{\alpha}{2} \right )\Gamma (a+D) \ \ \ \ \hbox{and} \ \ \ \
\Gamma(a+D,\lambda_{UB}(b+S)) = \Gamma (a+D) \ \frac{\alpha}{2},   \label{credi}
\end{equation}
respectively.
Here $\Gamma(a+D,x)$ is the incomplete gamma function and $\ds S = \sum_{i=1}^D x_{i:n} + (n-D)T$.  Observe that, when 
$a = b = 0$, the Bayes estimate of $\lambda$ matches with $\widehat{\lambda}$, although it is an improper prior.  Therefore, 
the comparison of the confidence intervals based on 
(\ref{conf-1}) and the credible interval based on (\ref{credi}) makes sense, although their interpretations
are different.  When $D$ = 0, the posterior density function of $\lambda$ becomes improper for $a = b$ = 0.  Due to this reason
we propose to use a proper prior with $a = b \approx$ 0 as suggested by Congdon (2014).

\section{\sc Numerical Comparisons}

In this section we present some simulation results to compare the performances of the two confidence intervals and the corresponding
credible intervals.  We compare the performances in terms of the average lengths and the coverage percentages.  Different values of 
$n$, $\lambda$ and $T$  are taken.  We consider $n$ = 5, 10, 15, 20, $\lambda$ = 0.5, 1.0, 2.0 and $T$ = 1, 2.  For Bayesian inference
we have taken $a = b$ = 0.001.  All the results are based on 5,000 replications.  The results are reported 
in Tables \ref{table-1} to \ref{table-3}.

The following points have been revealed from these simulation experiments.  In all these cases it is observed that  biases and  mean 
squared errors (MSEs) decrease as sample size increases, $T$ increases or $\lambda$ increases as expected.  Now comparing the confidence
intervals and credible intervals in Tables \ref{table-1} to \ref{table-3} it is 
clear that for small sample sizes, small $\lambda$ and small $T$ values, the confidence intervals based on unconditional distribution performs
better that the credible intervals based on non-informative priors in terms of the coverage percentages.  The coverage percentages 
of the confidence intervals based on unconditional distribution are very close to the nominal value (95\%) in all cases.  Although, as 
expected for large sample sizes they are almost equal.
The confidence intervals based on the conditional distribution do not perform very well when the sample size is very small and $\lambda$
is also small, although when the sample size is not very small it performs well.   Performances of the two confidence intervals match 
as expected when sample size is large.

\section{\sc Some Related Models}

In this section we consider some of the related two-parameter models, and construct an exact 100$(1-\alpha)\%$ confidence set of the two 
parameters, when $n$ items are tested and there is no observed failure during $[0, T]$.

\noindent Model 1: A two-parameter exponential distribution with the location parameter $\mu$ and scale parameter $\lambda$ 
having the following PDF
\begin{equation}
f(x; \lambda, \mu) =  \begin{cases}
\lambda e^{-\lambda(x-\mu)}, & \text{if $x \geq \mu$,}\\
0, & \text{if $x < \mu$}.
\end{cases}
\end{equation}
Here $\lambda > 0$ and $-\infty < \mu < \infty$.  In this case $P_{\{\lambda, \mu\}}(D = 0) = e^{-n \lambda (T-\mu)}$.  Hence, a 100$(1-\alpha)\%$ 
confidence set of $(\mu, \lambda)$ can be obtained as $A = \{(\mu, \lambda); \lambda > 0, -\infty < \mu < \infty, 
P_{\{\lambda, \mu\}}(D = 0) \ge (1-\alpha)\}$.  Therefore, 
$$
A = \{(\mu, \lambda); \lambda > 0, -\infty < \mu < T, \lambda (T-\mu) \le -\ln (1-\alpha)/n \}.
$$

\noindent Model 2:  Let us consider now a two-parameter Weibull distribution with the shape parameter $\beta$ and scale parameter 
$\lambda$ which has the following PDF

\begin{equation}
f(x; \lambda, \beta) = \begin{cases}
\beta \lambda x^{\beta-1} e^{-\lambda x^{\beta}}, & \text{if $ x \geq 0$},\\
0, & \text{if $x<0$}.
\end{cases}
\end{equation}
Here $\beta > 0$ and $\lambda > 0$.  In this case $P_{\{\lambda, \beta\}}(D = 0) = e^{-n \lambda T^{\beta}}$.  A 100$(1-\alpha)\%$ confidence
set of $(\lambda, \beta)$, can be obtained as $A = \{(\lambda, \beta); \lambda > 0, \beta > 0, 
P_{\{\lambda, \beta\}}(D = 0) \ge (1-\alpha)\}$.  Hence, 
$$
A = \{(\lambda, \beta); \lambda > 0, \beta > 0, \lambda T^{\beta} \le -\ln (1-\alpha)/n \}.
$$

\noindent Model 3: Similarly, if we consider two-parameter generalized exponential distribution which has the following PDF
\begin{equation}
f(x; \lambda, \beta) = \begin{cases}
\beta \lambda e^{-\lambda x} (1 - e^{-\lambda x})^{\beta-1}, & \text{if $x \geq 0$},\\
0, & \text{if $x<0$}.
\end{cases}
\end{equation}
Here $\beta > 0$ is the shape parameter and $\lambda > 0$ is the scale parameter.  Here, $P_{\{\lambda, \beta\}}(D = 0) = (1 - (1 - e^{-\lambda T})^\beta)^n$.  A 100$(1-\alpha)\%$ confidence
set of $(\lambda, \beta)$, can be obtained as $A = \{(\lambda, \beta); \lambda > 0, \beta > 0, 
P_{\{\lambda, \beta\}}(D = 0) \ge (1-\alpha)\}$.  Therefore,
$$
A = \{(\lambda, \beta); \lambda > 0, \beta > 0, (1 - e^{-\lambda T})^{\beta} \le 1-(1-\alpha)^{1/n}\}.
$$

\noindent {\sc Joint Confidence Set}

\begin{figure}[H]
\centering
\begin{subfigure}{0.32\textwidth}
\centering
\includegraphics[width = \textwidth]{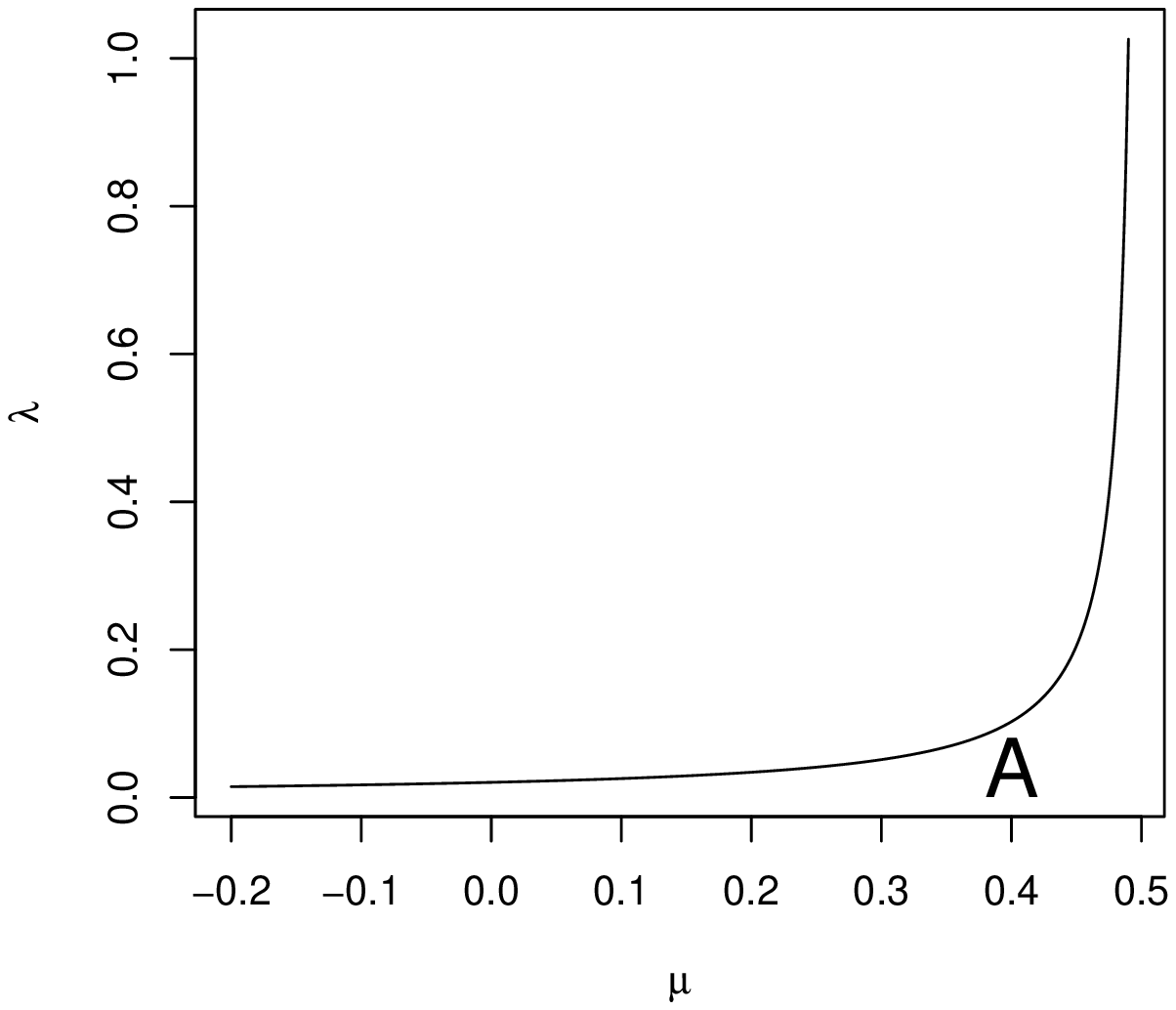}
\caption{Joint confidence set of $(\lambda, \mu)$ for the two-parameter exponential distribution.   \label{jcs-exp}}
\label{fig:left}
\end{subfigure}
\begin{subfigure}{0.32\textwidth}
\centering
\includegraphics[width = \textwidth]{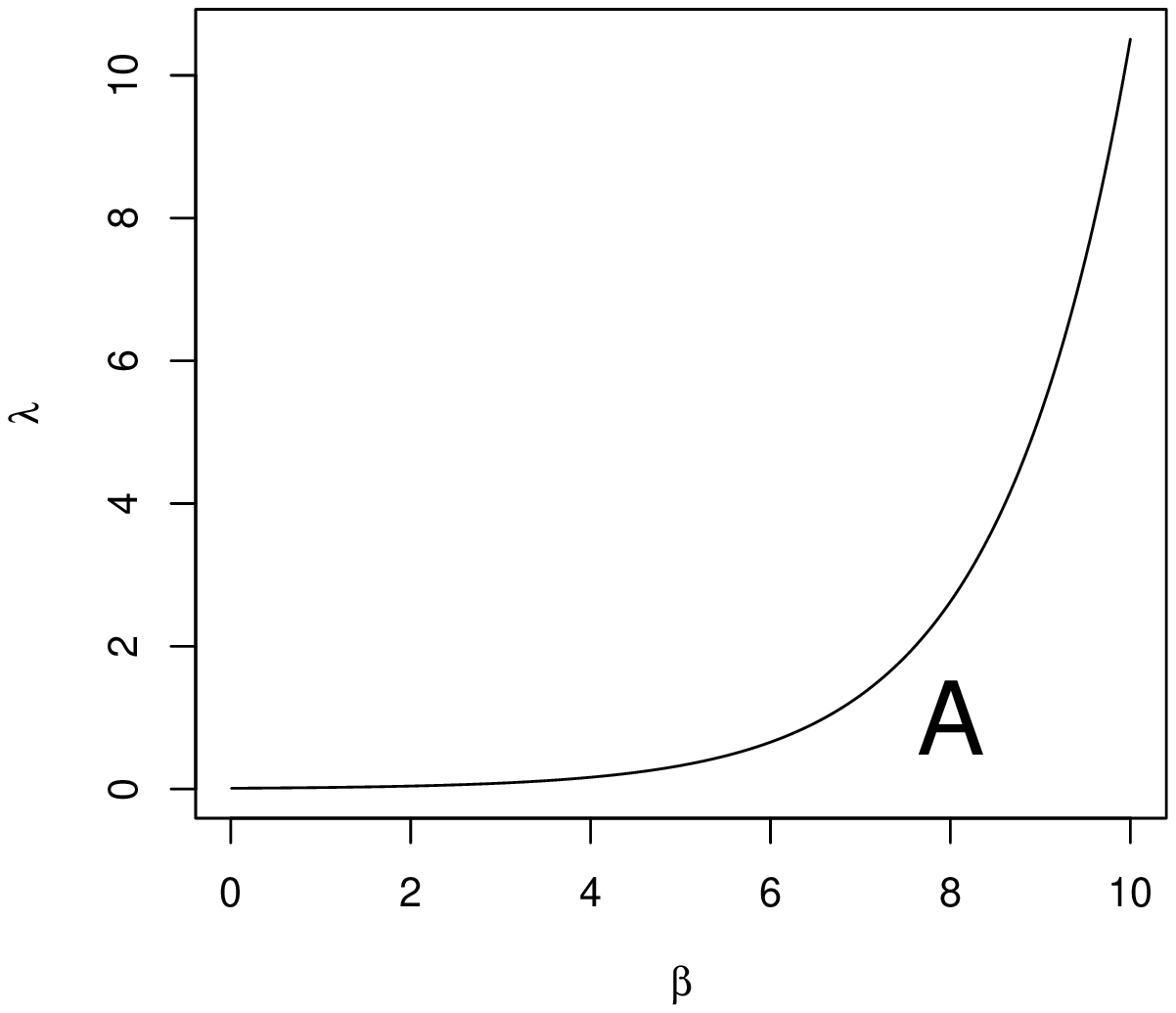}
\caption{Joint confidence set of $(\beta, \lambda)$ for the two-parameter Weibull distribution.    \label{jcs-web}}
\label{fig:right}
\end{subfigure}
\begin{subfigure}{0.32\textwidth}
\centering
\includegraphics[width = \textwidth]{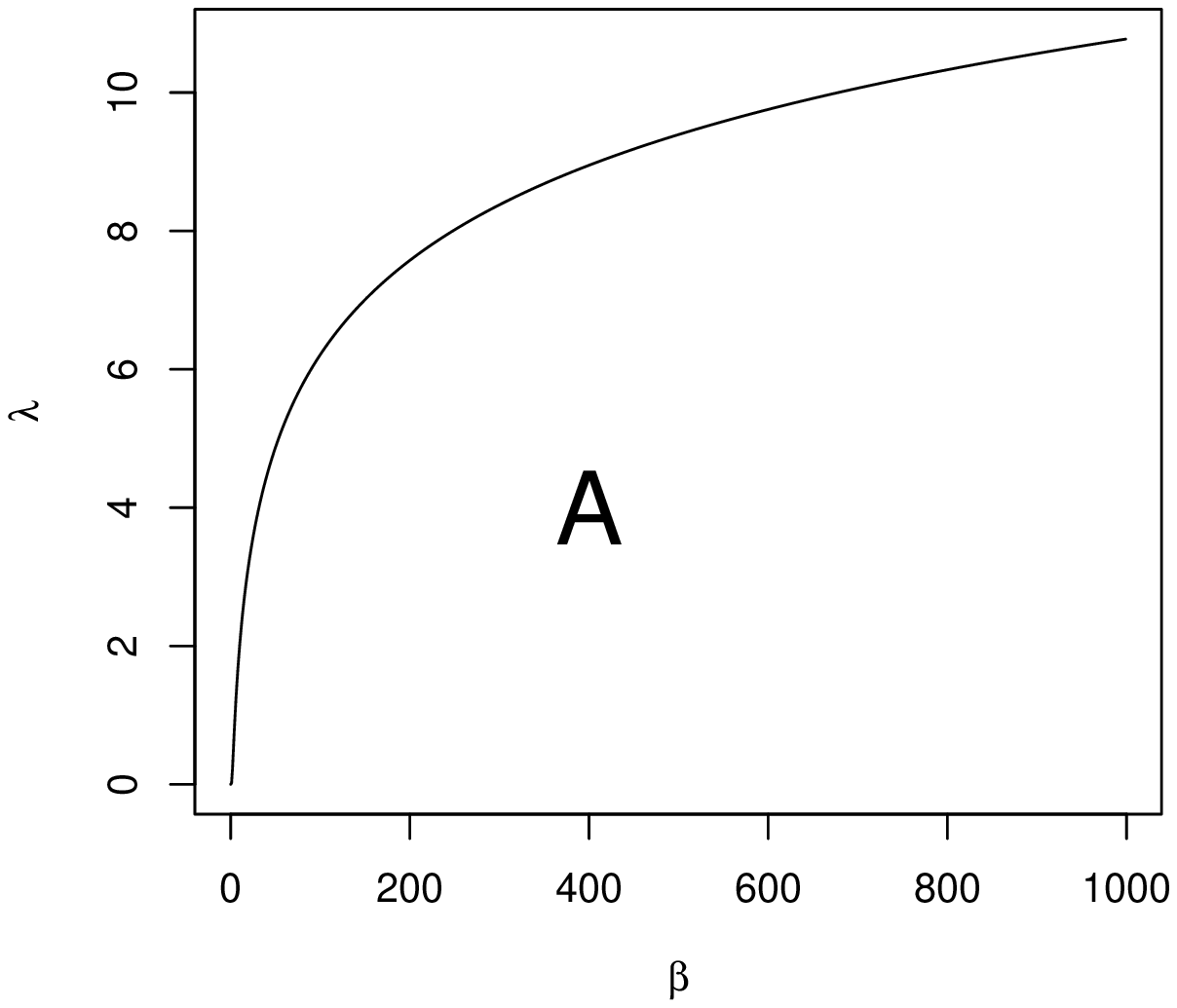}
\caption{Joint confidence set of $(\beta, \lambda)$ for the two-parameter generalized exponential distribution.   \label{jcs-ge}}
\label{fig:right}
\end{subfigure}
\caption{Joint confidence sets of the parameters for different distributions.  Here `{\bf A}' denotes the required set. }
\label{fig:combined}
\end{figure}

Just for illustrative purposes, taking $n$ = 5 and $T$ = 0.5, we provide the joint 100$(1-\alpha)\%$ 
confidence set as the shaded region of (i) $(\lambda, \mu)$ 
for two-parameter exponential distribution, (ii) $(\beta,\lambda)$ for two-parameter Weibull distribution and (iii) $(\beta, \lambda)$ for two-parameter generalized exponential distribution, in Figure \ref{fig:combined} when
$D$ = 0.


\section{\sc Conclusions}

In this paper we have considered the time truncated exponential distribution and provide the exact confidence interval of the 
unknown scale parameter based on the unconditional argument.  All the existing results are based on the conditional argument and 
it does not provide any statistical inference of the unknown parameter when there is no observation.  In this paper we have provided
the inference on the scale parameter even when there is no observation.  Simulation experiments are performed and it is observed that 
the proposed method works quite well even when the sample size is very small.  The joint confidence sets have been provided for 
two-parameter exponential, Weibull and generalized exponential distributions also for time truncated case when there is no 
observation during the time period of the experiment.  The results can be extended to the competing risks model also, as considered in 
Kundu, Kannan and Balakrishnan (2004).



\begin{table}
\caption{CI of $\lambda$ based on unconditional distribution of $\widehat{\lambda}$ when $T$=1, 2  \label{table-1}}
\vspace{0.15in}
\centering
\begin{tabular}{|c| c | c c c c |c c c c|}
\hline
  & 	& 	& T=1 &   &  & 	& T=2 &   &  \\
\hline
  & $n$	& Bias	& MSE &  CI & CP & Bias	& MSE &  CI & CP \\  \hline
 & 5 &	0.074	& 0.207	& 1.577	& 97.36	& 0.091 &	0.166	& 1.253	& 94.42 \\
 & 10&	0.033	& 0.080	& 1.041	& 94.10	& 0.038 &	0.055	& 0.825	& 94.62 \\
$\lambda$=0.5 &	15 & 0.020 & 0.049 & 0.833 & 94.86 & 0.023	& 0.033	& 0.659	& 94.58\\
 & 20	& 0.015	& 0.035	& 0.716	& 94.86	& 0.016	& 0.023	& 0.565	& 94.96 \\
\hline
 & 5 &	0.182	& 0.662	& 2.505	& 94.42	& 0.226	& 0.619	& 2.240	& 94.50	 \\
 & 10 &	0.076	& 0.221	& 1.649	& 94.62	&  0.092	& 0.181	& 1.434	& 94.60\\
$\lambda$=1.0	& 15	& 0.046	& 0.132	& 1.318	& 94.58	& 0.056	& 0.104	& 1.139	& 94.56\\
 & 20	& 0.032	& 0.092	& 1.130	& 94.98	& 0.040	& 0.071	& 0.974	& 94.86\\
 	\hline
& 5 &	0.452	& 2.477	& 4.481	& 94.50	& 0.510 	& 2.428	& 4.351	& 94.66\\
 & 10	& 0.183	& 0.724	& 2.869	& 94.60	& 0.214	& 0.688	& 2.742	& 94.42\\
$\lambda$=2.0	& 15 & 0.112 &	0.416	& 2.278	& 94.56	& 0.131	& 0.390	& 2.162	& 94.92	\\
 & 20	& 0.081	& 0.284	& 1.948	& 94.86	& 0.094	& 0.260	& 1.843	& 94.92\\
	\hline
\end{tabular}
\end{table}

\begin{table}
\caption{CI of $\lambda$ based on the conditional distribution of $\widehat{\lambda}$ when $T$=1, 2   \label{table-2}}
\vspace{0.15in}
\centering
\begin{tabular}{|c c | c c c c | c c c c|}
\hline
 & 	& 	& T=1 &   &  & 	& T=2 &   &  \\
\hline
 & $n$	& Bias	& MSE &	CI & CP & Bias	& MSE &	CI & CP \\ \hline
 & 5	& 0.067	& 0.185	& 1.398	& 69.28	& 0.088 & 0.156 & 1.241 & 91.28\\ 
 &10	& 0.034	& 0.072	& 1.037	& 92.74	& 0.038 & 0.055 & 0.826 & 94.62\\ 
$\lambda$=0.5	& 15	& 0.021	& 0.048	& 0.837	& 94.96	& 0.023 & 0.033 & 0.659 & 94.58\\ 
 & 20	& 0.016	& 0.035	& 0.718	&  94.88 & 0.016 &	0.023 &	0.565 &	94.96	\\ 	\hline
 & 5	& 0.176	& 0.623	& 2.482	& 91.28	& 0.226 &	0.618 &	2.245 &	94.48\\ 
 & 10	& 0.076	& 0.219	& 1.652	& 94.62	& 0.092 &	0.181 &	1.435 &	94.60\\ 
$\lambda$=1.0	& 15	& 0.046	& 0.132	& 1.318	& 94.58	& 0.056 & 0.104 & 1.139 & 94.56\\ 
 & 20	& 0.032	& 0.092	& 1.130	& 94.98	& 0.040 &	0.071 &	0.974 &	94.86\\ 	\hline	
 & 5	& 0.452	& 2.471	& 4.490	& 94.48	& 0.510 & 2.428 &	4.352 &	94.66 \\ 
 & 10	& 0.184	& 0.724	& 2.869	& 94.60	& 0.214 &	0.688 &	2.742 &	94.42\\ 
$\lambda$=2.0	& 15 &	0.112	& 0.416	& 2.278	& 94.56	& 0.131 & 0.390 & 2.162 & 94.92\\ 
 & 20	& 0.081	& 0.284	& 1.948	& 94.86	& 0.094 &	0.260 &	1.843 &	94.92\\ 	\hline
\end{tabular}
\end{table}

\begin{table}
\caption{CRI of $\lambda$ based on non-informative priors when $T$=1, 2   \label{table-3}}
\vspace{0.15in}
\centering
\begin{tabular}{|c c | c c c c|c c c c|}
\hline
 & 	& 	& T=1 &   &  & 	& T=2 &   &  \\
 \hline
 & $n$	& Bias	& MSE	& CRI &	CP & Bias	& MSE	& CRI &	CP\\
\hline 
 & 5 & 0.074 & 0.207 & 1.390 &	89.36 & 0.091	& 0.166	& 1.202	&	90.42\\
 &10 & 0.033 & 0.080 & 0.985 &	92.36 & 0.038	& 0.055	& 0.807	&	93.78\\
$\lambda$=0.5 &	15 & 0.020 & 0.049 & 0.804 & 94.94 &  0.023	& 0.033	& 0.649	& 93.76\\
 & 20 &	0.015 &	0.035 &  0.697	& 93.28 & 0.016	& 0.023	& 0.559	&	94.28 \\
\hline
 & 5 &	0.182 &	0.662 &	2.404 &	90.42 & 0.226	& 0.619	& 2.212	&	93.54\\
 & 10 &	0.076	& 0.221	& 1.614	& 93.78 & 0.092	& 0.181	& 1.424	&	94.04 \\
$\lambda$=1.0 &	15 & 0.046 & 0.132 & 1.299 & 93.76 &  0.056	& 0.104	& 1.134	& 94.04\\
 & 20 &	0.032 &	0.092 &	1.117 &	94.28 & 0.040	& 0.071	& 0.970	& 94.58 \\
	\hline
 & 5 & 0.452 & 2.477 & 4.422 & 93.54 & 0.510	& 2.428	& 4.343	& 94.46\\
 & 10 &	0.183 &	0.724 &	2.849 &	94.04 & 0.214	& 0.688	& 2.738	& 94.26 \\
$\lambda$=2.0 &	15 & 0.112 & 0.416 & 2.267 & 94.06 & 0.131 &	0.390	& 2.161	&	94.76\\
 & 20 &	0.081 &	0.284 &	1.941 &	94.58 & 0.094	& 0.260	& 1.842	&	94.80\\
	\hline
\end{tabular}
\end{table}

\end{document}